\documentstyle[aps,prb]{revtex}
\begin{document}
\draft
\title{Monte-Carlo simulation of the coherent backscattering of electrons in a
ballistic system}
\author{K. L. Janssens and F. M. Peeters\cite{email}}
\address{Department of Physics, University of Antwerp (UIA),\\
B-2610 Antwerpen, Belgium}
\date{\today}
\maketitle

\begin{abstract}
We study weak localization effects in the ballistic regime as induced by
man-made scatterers. Specular reflection of the electrons off these
scatterers results into backscattered trajectories which interfere with
their time-reversed path resulting in weak localization corrections to the
resistance. Using a semi-classical theory, we calculate the change in
resistance due to these backscattered trajectories. We found that the
inclusion of the exact shape of the scatterers is very important in order to
explain the experimental results of Katine {\it et al.} [Superlattices and
Microstructures {\bf 20}, 337 (1996)].
\end{abstract}


Recently, Katine {\it et al.} \cite{katine} observed coherent backscattering
in a high mobility two dimensional electron gas (2DEG), containing a quantum
point contact and two reflector gates (see Fig. 1). The experiment was
performed in the ballistic regime. The backscattered trajectories lead to an
increase of the resistance which can be removed by the application of a
perpendicular magnetic field like in the well-known weak localization effect 
\cite{alts1}. The aim of the present work is to explain this experimental
result. In contrast to earlier measurements of coherent backscattering, many
backscattered trajectories are found in this system, which inclose almost
all the same surface area. The constructive interference of these
backscattered trajectories leads to an enhancement of the resistance, which
is known as {\it weak localization}. Previously, weak localization was
always measured in the diffuse regime, where it is a consequence of the
scattering on impurities. In this work we assume a ballistic regime, where
the scattering on impurities is negligeable and the reflector gates will now
act as big man-made impurities.

For the description of the electron motion we use the semi-classical
billiard model \cite{beenakker} in which electrons are point particles. Note
that the dimensions of the system under consideration (2$\mu m$) are larger
than the Fermi wavelength ($\lambda _{F}=458$\AA , $n=3\times 10^{15}m^{-2}$
in the experiment). Only electrons on the Fermi surface are important for
conduction and therefore the relevant electrons all have the same velocity $%
v_{F}=${\it 
%TCIMACRO{\UNICODE{0x127}}%
%BeginExpansion
h\hskip-.2em\llap{\protect\rule[1.1ex]{.325em}{.1ex}}\hskip.2em%
%EndExpansion
}$k_{F}/m=2\times 10^{5}m/s$ but they are injected from the point contact
under an angle $\alpha $ which satisfies the distribution $P(\alpha )=\cos
(\alpha )/2.$ The electron trajectories are determined by Newtons law which,
in the absence of a magnetic field, leads to motion in a straight line,
while the electrons describe a circular trajectory in the presence of a
magnetic field \cite{peeters}. A condition for the applicability of the
ballistic billiard model is that there is no scattering on impurities and
that $\lambda _{F}$ is much smaller than the size of the reflectors. In the
experiment the mean free path was $l_{e}=5.4\mu m$ which is substantially
larger than 2$\mu m,$ the dimensions of the system. The scattering off the
reflector gates is specular. This results in a resistance $R,$ determined by
the Landauer formula \cite{landauer} 
\begin{equation}
R=h/2e^{2}N(1-r)\text{ .}  \label{landauer}
\end{equation}
Hereby $h/e^{2}=25%
%TCIMACRO{\unit{k\UNICODE{0x3a9}}}%
%BeginExpansion
\mathop{\rm k%
\Omega}%
%EndExpansion
$ represents the fundamental unit of resistance, $N$ is the number of
conducting channels and $r$ is the probability for backscattering which is
calculated using our numerical simulation.

The Monte Carlo program calculates the trajectory of the electron and
verifies whether the electron returns to the point contact or not. The
relative number of returning electrons determines $r.$ About 10 million
electrons were injected from the point contact (which was taken 0.17$\mu m$
wide) at each value of the magnetic field. The electron flow is shown in
Fig. 2 for $B=0T.$ From the Landauer formula \ref{landauer} the resistance $%
R $ due to the returning electron orbits\ is obtained. The total resistance
as measured experimentally equals this resistance $R$ plus the point contact
resistance $R_{pc}$ and the resistance due to the two-dimensional electron
gas (2DEG). The theoretical result for $R$ is shown in Fig. 3. For $\left|
B\right| <20mT$ the resistance increases with increasing magnetic field
which is opposite to what was found experimentally (see solid curve in Fig.
4). Theoretically, the increase in resistance is a consequence of the fact
that the electrons ejected from the point contact have, due to the $P(\alpha
)=\cos (\alpha )/2$ distribution, a higher probability to be ejected into
the forward direction. With increasing magnetic field, this beam is steered
towards one of the reflectors, resulting in an increase of the number of
electrons returning to the point contact.

The discrepancy in the functional behaviour of $R$ versus $B$ between theory
and experiment demonstrates clearly that the wave nature of the electrons
has to be included. We will include this by taking into account the phase of
the electron. The quantum mechanical contribution to the resistance is
determined by the interference of the electron trajectories with their
time-reversed paths. The probability for a particle to return to the point
contact after collision with the reflector gates is given by $P=\left|
\sum\nolimits_{j}A_{j}\right| ^{2}$, where $A_{j}=\left| A_{j}\right|
e^{i\theta _{j}}$ is the amplitude of the returning trajectory $j$. The
change of phase $\theta _{j}$ is determined by the magnetic flux through the
surface area $S_{j}$ inclosed by the trajectory of the electron, which is
given by $\theta _{j}=eBS_{j}/${\it 
%TCIMACRO{\UNICODE{0x127}}%
%BeginExpansion
h\hskip-.2em\llap{\protect\rule[1.1ex]{.325em}{.1ex}}\hskip.2em%
%EndExpansion
}. For a time-reversed path, this change of phase has an opposite sign. When
two trajectories $A_{j}$ en $A_{k}$ as in Fig. 1 are considered, we have $%
\theta _{j}\simeq -\theta _{k}$ and $\left| A_{j}\right| \simeq \left|
A_{k}\right| .$ This implies 
\begin{equation}
P=\left| 
%TCIMACRO{\underset{j}{\sum }}%
%BeginExpansion
\mathrel{\mathop{\sum }\limits_{j}}%
%EndExpansion
2\left| A_{j}\right| \cos \theta _{j}\right| ^{2}\text{ .}
\label{waarschijnlijkheid}
\end{equation}
First we checked that for every returning trajectory with $\alpha _{i}>0$
there exists an equivalent time-reversed path which has $\alpha _{j}<0$.
This is illustrated in Fig. 5 where we show the percentage of returning
trajectories as function of the angle under which the electron is ejected
from the point contact for two different magnetic fields: (a) $B=0T$ and (b) 
$B=10mT.$ Notice\ that the magnetic field results in: 1) a shift of the
starting angle of the returning trajectories, and 2) in a small imbalance of
the returning trajectories with $\alpha _{i}>0$ and $\alpha _{i}<0.$ But we
found that the latter has almost no effect on the resistance, and therefore
will be ignored.

Next we used Eq. (\ref{waarschijnlijkheid}) in order to calculate $r=P/$%
(number of simulated electrons), which we inserted in the Landauer formula
(Eq. \ref{landauer}). The obtained result is shown in Fig. 4 as the dotted
curve. Notice that the positions of the peaks are well reproduced but the
height of the central peak is substantially underestimated. The oscillations
are determined by the phase difference between the returning trajectories
and their time-reversed paths. Constructive interference occurs for $eBS/$%
{\it 
%TCIMACRO{\UNICODE{0x127}}%
%BeginExpansion
h\hskip-.2em\llap{\protect\rule[1.1ex]{.325em}{.1ex}}\hskip.2em%
%EndExpansion
}$=2\pi n$ (for $n=1,2,...$) and there is a higher probability for returning
to the starting point. This results in a peak in the resistance which occurs
with a period $\Delta B=h/2eS_{0}=3.33mT,$ where we took $S_{0}=6.21\times
10^{-13}m^{2}$. For destructive interference, i.e. $eBS/${\it 
%TCIMACRO{\UNICODE{0x127}}%
%BeginExpansion
h\hskip-.2em\llap{\protect\rule[1.1ex]{.325em}{.1ex}}\hskip.2em%
%EndExpansion
}$=\pi (2n+1)$ (for $n=1,2,...$), the probability for returning to the
starting point is smaller which determines the position of the dip.

In Ref. \cite{katine} a simplified model was proposed which could explain
the experimental results. This model assumed a gaussian distribution of
surface areas of the backscattered trajectories with average $%
S_{0}=6.21\times 10^{-13}m^{2}$ and standard deviation $\sigma $ which was
taken as a fitting parameter where $\sigma =0.5S_{0}$ was found. Using our
billiard model we calculated the distribution of the surfaces of the
returning trajectories. The result is shown in Fig. 6 which can be fitted to
a Gaussian (solid curve) with average $S_{0}$ and width $\sigma .$ For $B=0T$
we found $S_{0}=6.74\times 10^{-13}m^{2}$ and $\sigma =0.68\times
10^{-13}m^{2}$ and for $B=10mT$ we found $S_{0}=6.43\times 10^{-13}m^{2}$
and $\sigma =1.13\times 10^{-13}m^{2}.$ This corresponds to respectively $%
\sigma \simeq 0.1S_{0}$ and $\sigma \simeq 0.17S_{0}$ which is a factor 4
smaller than used in Ref. 1. Therefore, we are forced to conclude that the
simple model of Ref. 1 is inadequate to explain their experimental results
because a smaller $\sigma $ will lead to a much smaller damping of the
oscillations in $R$ as a function of the magnetic field.

In order to explain the existing discrepancy between our theoretical result
and the experiment we investigated the effect of the shape of the reflectors
on the number of returning electrons. In the above analysis we assumed that
the reflector gates are straight lines. However in reality they are part of
parabolas given by \cite{katine} 
\begin{eqnarray*}
x &=&f(\theta )\cos \theta  \\
y &=&f(\theta )\sin \theta  \\
f(\theta ) &=&\frac{2r_{0}}{1+\sin \theta }\text{ \ \ \ \ \ where \ \ 12}%
%TCIMACRO{\UNICODE[m]{0xb0}}%
%BeginExpansion
{{}^\circ}%
%EndExpansion
<\theta <30%
%TCIMACRO{\UNICODE[m]{0xb0}}%
%BeginExpansion
{{}^\circ}%
%EndExpansion
\text{ .}
\end{eqnarray*}
At first sight our previous assumption seemed very reasonable because these
parabolas are very well approximated by straight lines. Nevertheless,
further study learns that the backscattering is much better with parabolic
reflector gates, because the opening of the point contact is right in the
focus of the parabolas. An electron, starting in this point and directed
towards one of the reflectors, is scattered back to exact the same point by
the parabolic reflector gates in the absence of a magnetic field. This is
clearly illustrated in the current flow diagram of Fig. 6 when compared with
Fig. 2. An applied magnetic field will much more strongly defocus the system
resulting in a smaller number of backscattered electrons as compared to the $%
B=0T$ situation. The resistance obtained with the parabolic reflectors is
shown in Fig. 4 by the dashed curve. The central peak and the second
neighbour peaks are nicely reproduced. The height of the first side peak is
still overestimated which may be due to the small influence of impurities on
the backscattering which was neglected in the present work.

In summary we studied the weak localization effect in a ballistic 2DEG with
man-made scatterers, i.e. parabolic reflectors, by means of computer
simulations. In the studied system the electrons could, after passing
through the point contact, collide with the reflector gates and be focussed
back to the point contact. The influence of this backscattering on the
magnetoresistance was studied as a function of the applied magnetic field.
We found that a pure classical treatment could not explain the experimental
results of Katine {\it et al. }and in fact has even the wrong magnetic field
dependence for the resistance. This was a clear indication that quantum
mechanical effects are in play. The quantum mechanical contribution was
included through the interference of the time-reversed paths. We found an
enhancement of the magnetoresistance as a consequence of the constructive
interference of backscattered electrons with their time-reversed paths. This
interference leads to oscillations in the resistance as function of the
magnetic field which are nicely reproduced by our theoretical results. In
order to find the correct qualitative behaviour of these oscillations, it
turned out to be necessary to include the correct shape (i.e. parabolic
reflector gates) of the reflectors in our simulation.

{\bf Acknowledgements.} This work was supported by IUAP-IV, the
Inter-University Micro-Electronics Center (IMEC, Leuven), and the Flemish
Science Foundation (FWO-Vl).

\begin{figure}[tbp]
\caption{Schematic view of the experimental system and an example of the
trajectory of a returning electron and its time-reversed path.}
\label{fig1}
\end{figure}

\begin{figure}[tbp]
\caption{The electron flow out of the point contact (situated at $x=0,$ $%
\left| y\right| <0.17\protect\mu m$) through the studied system at $B=0T.$}
\label{fig2}
\end{figure}

\begin{figure}[tbp]
\caption{The magnetoresistance as function of the applied magnetic field, as
obtained within a pure classical treatment.}
\label{fig3}
\end{figure}

\begin{figure}[tbp]
\caption{The experimental result (solid curve) for the resistance as
function of the magnetic field, together with the theoretical traces for
straight (dotted curve) and parabolic (dashed curve) reflector gates.}
\label{fig4}
\end{figure}

\begin{figure}[tbp]
\caption{The percentage of returning electrons as function of the angle at
which the electrons are ejected from the point contact for (a) $B=0T$ and
(b) $B=10mT.$}
\label{fig5}
\end{figure}

\begin{figure}[tbp]
\caption{The distribution of the inclosed surface areas of the backscattered
trajectories for (a) $B=0T$ and (b) $B=10mT$. The solid curve is a Gaussian
fit to the results obtained from our simulation.}
\label{fig6}
\end{figure}

\begin{figure}[tbp]
\caption{The same as Fig. 2 but now for parabolic reflector gates.}
\label{fig7}
\end{figure}


\begin{references}
\bibitem[{*}]{email}  Electronic address: peeters@uia.ua.ac.be

\bibitem{katine}  J. A. Katine, M. A. Eriksson, R. M. Westervelt, K. L.
Campman, and A. C. Gossard, Superlattices and Microstructures {\bf 20}, 337
(1996).

\bibitem{alts1}  B. L. Altshuler and P. A. Lee, Physics Today {\bf 41},
December 1988, p. 36.

\bibitem{beenakker}  C. W. J. Beenakker and H. van Houten, Phys. Rev. Lett. 
{\bf 63}, 1857 (1989).

\bibitem{peeters}  F. M. Peeters and X. Q. Li, Appl. Phys. Lett. {\bf 72},
572 (1998).

\bibitem{landauer}  R. Landauer, IBM J. Res. Dev. {\bf 1}, 223 (1957).
\end{references}
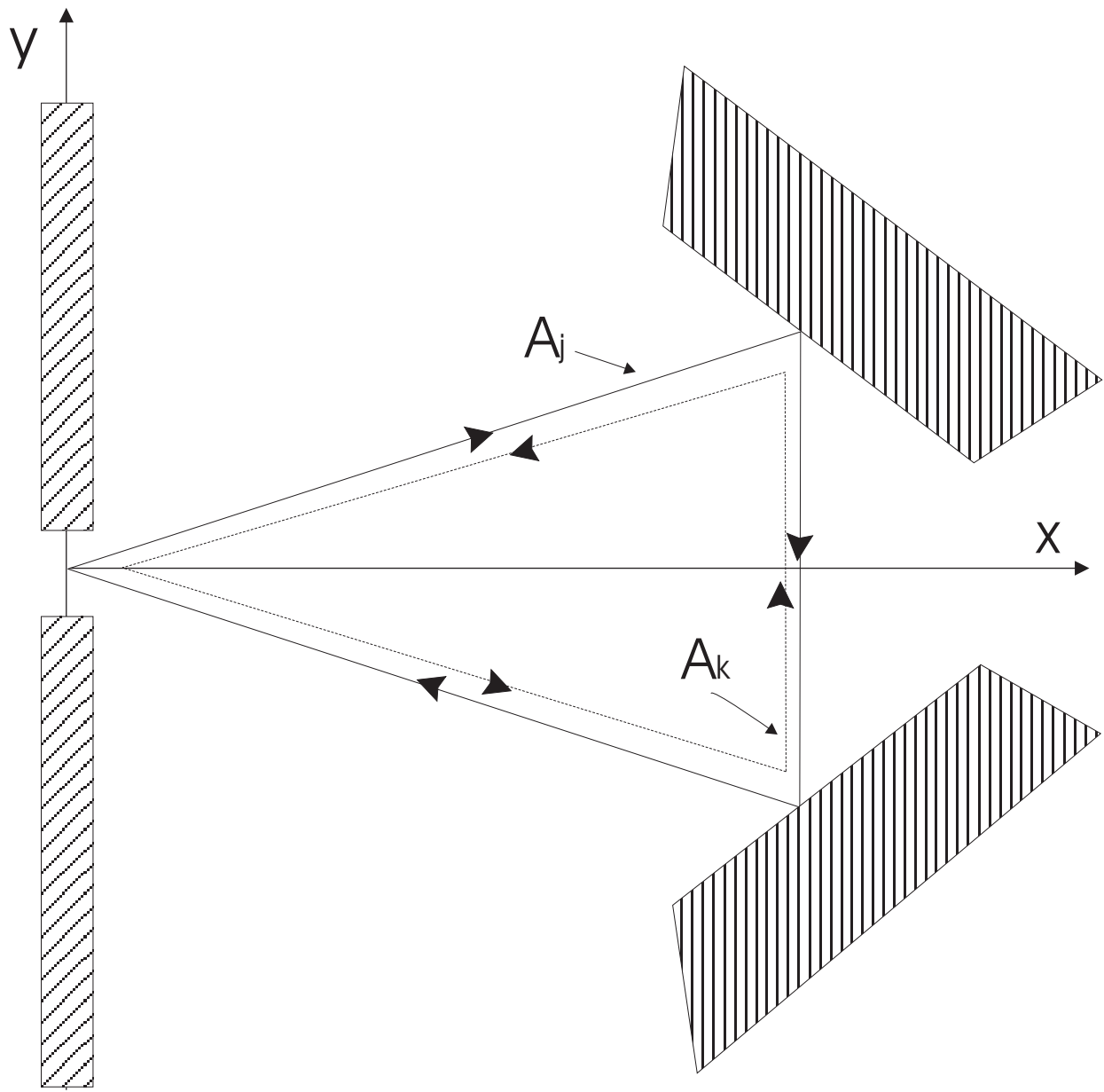
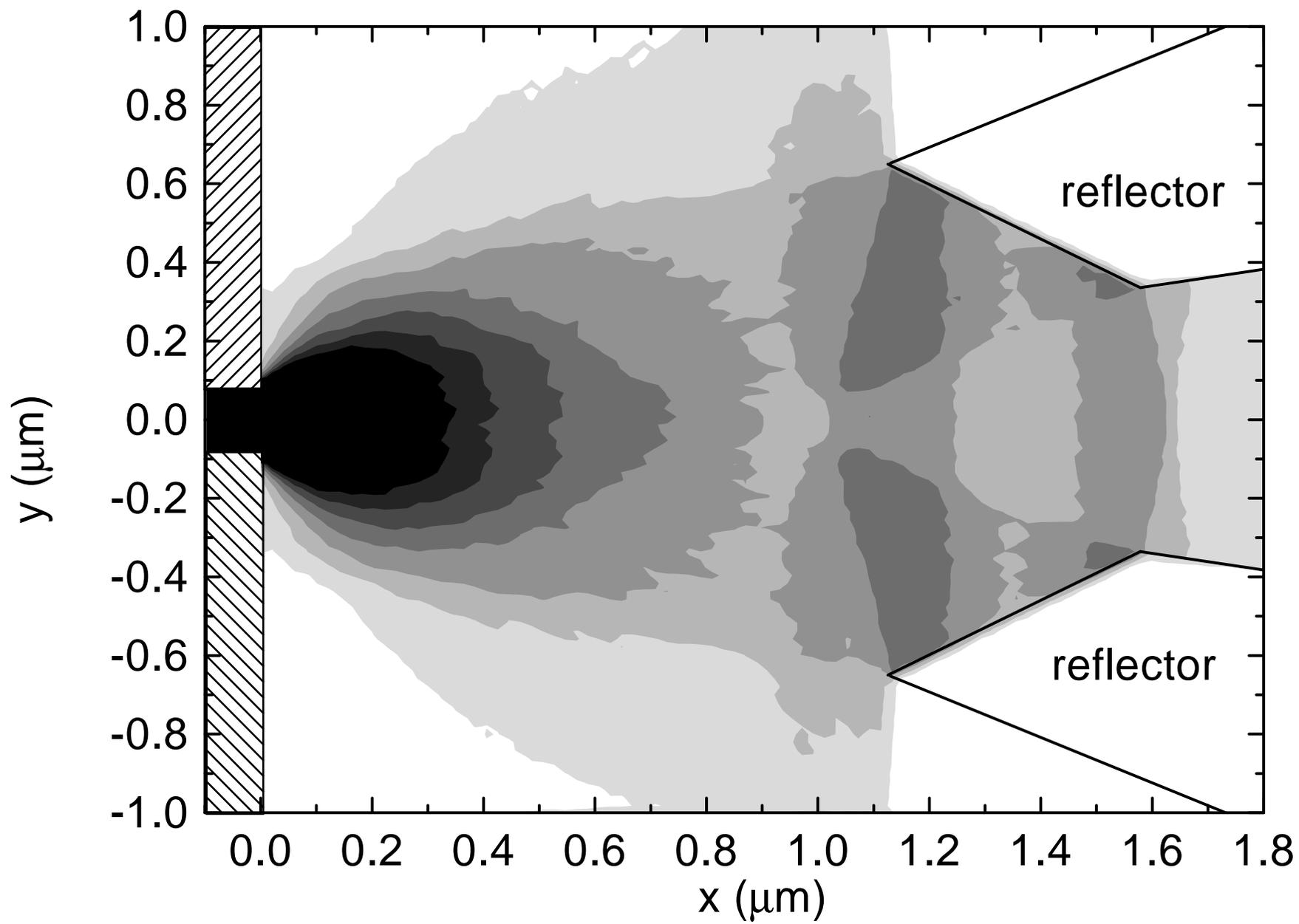
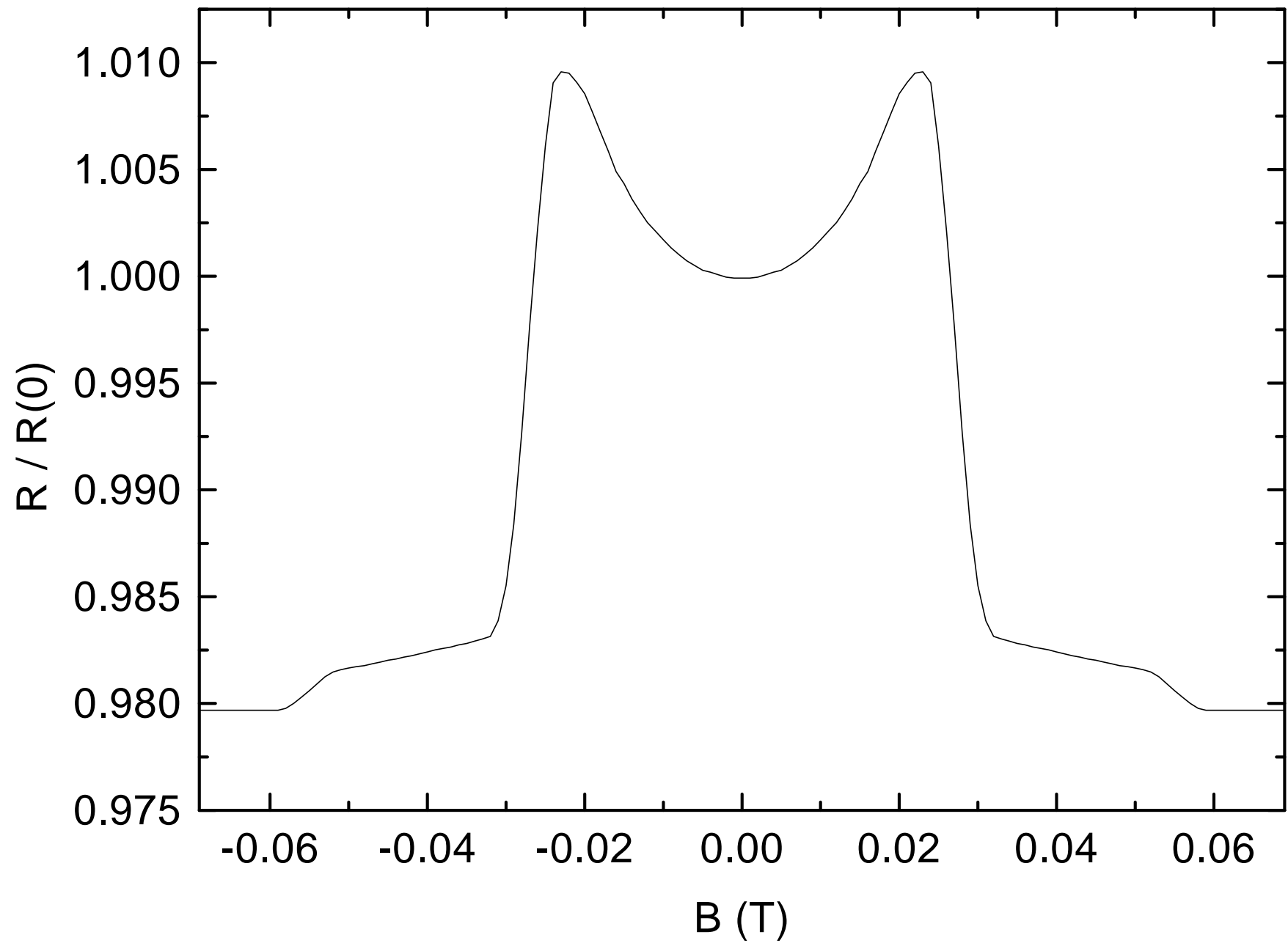
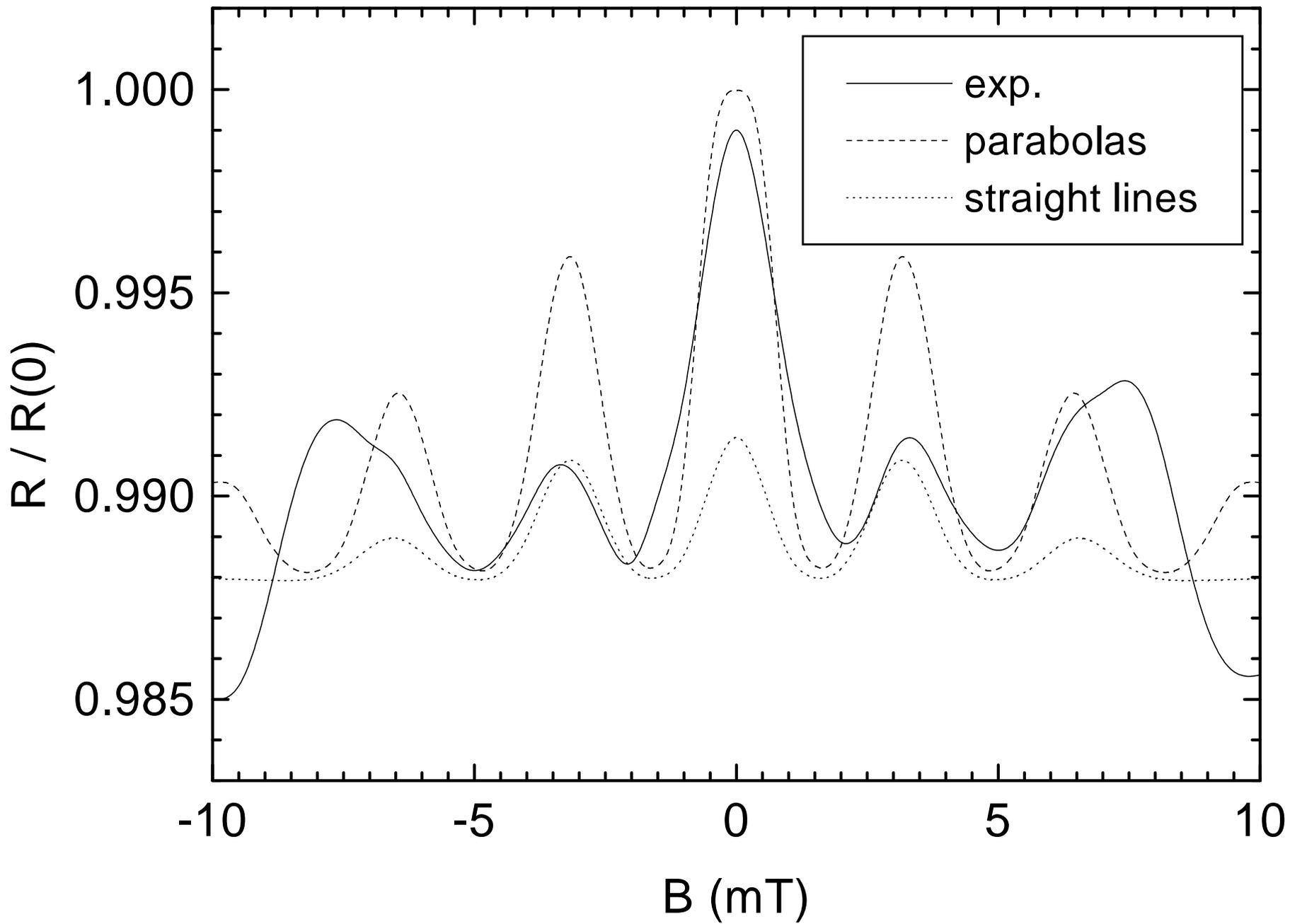
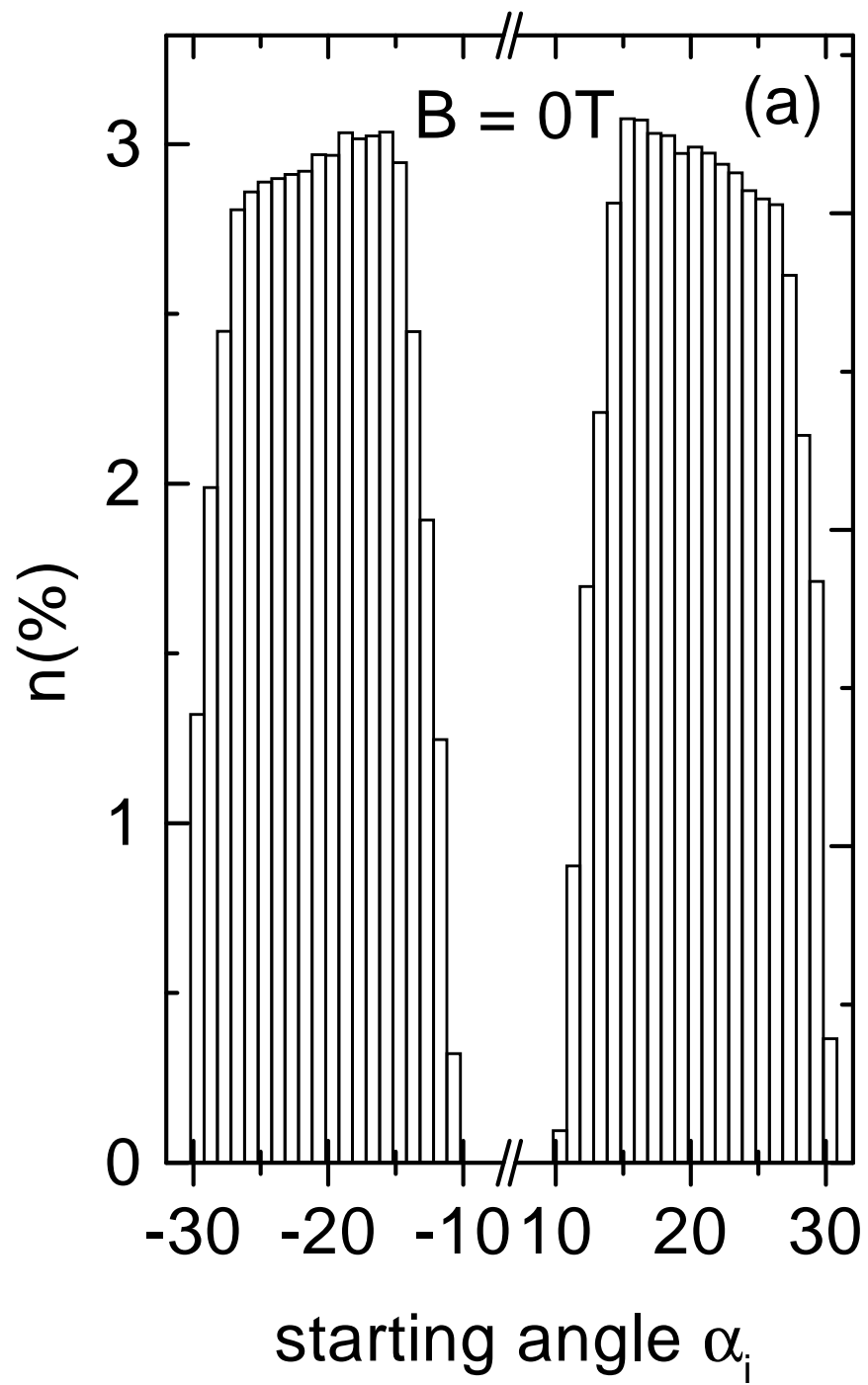
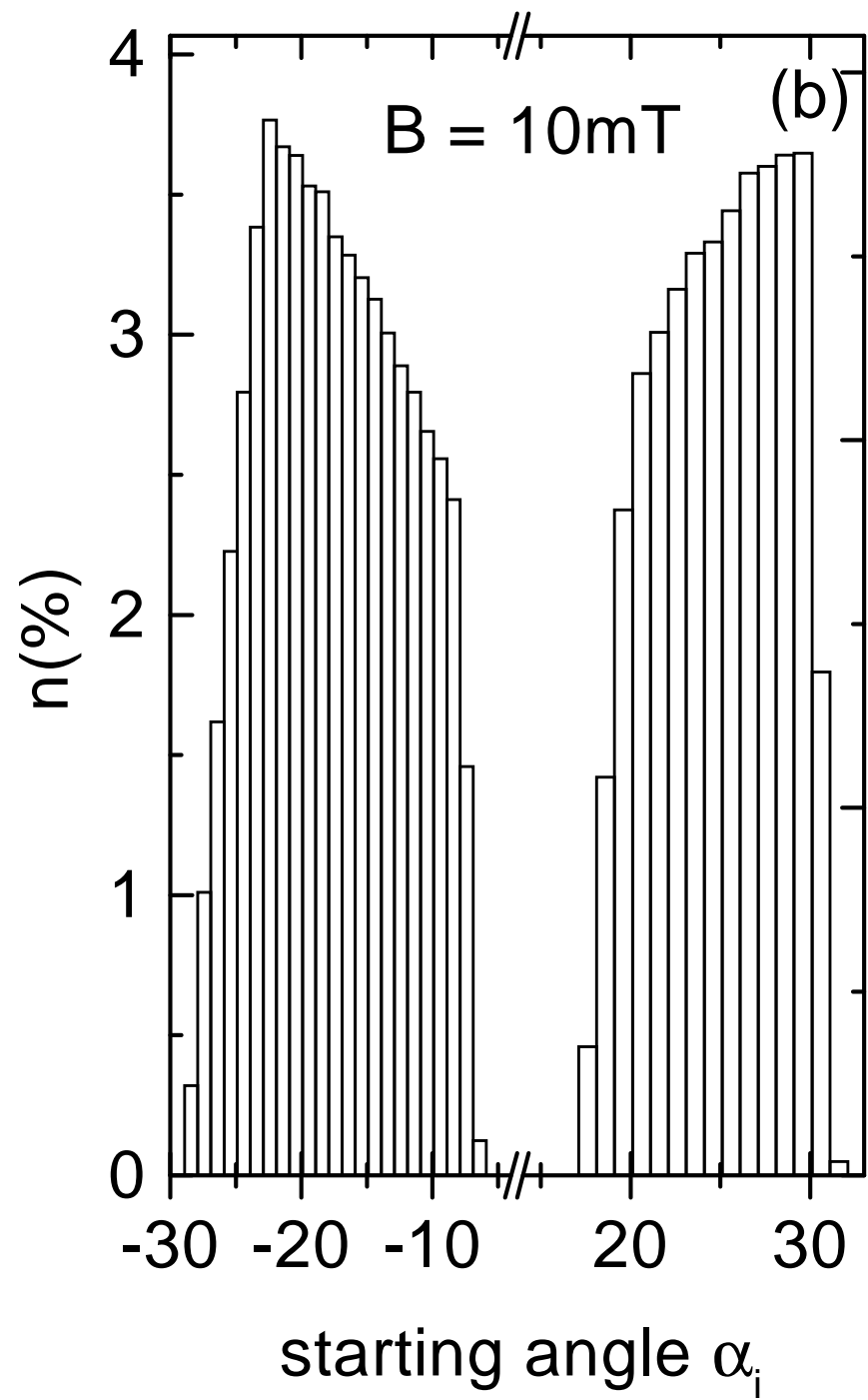
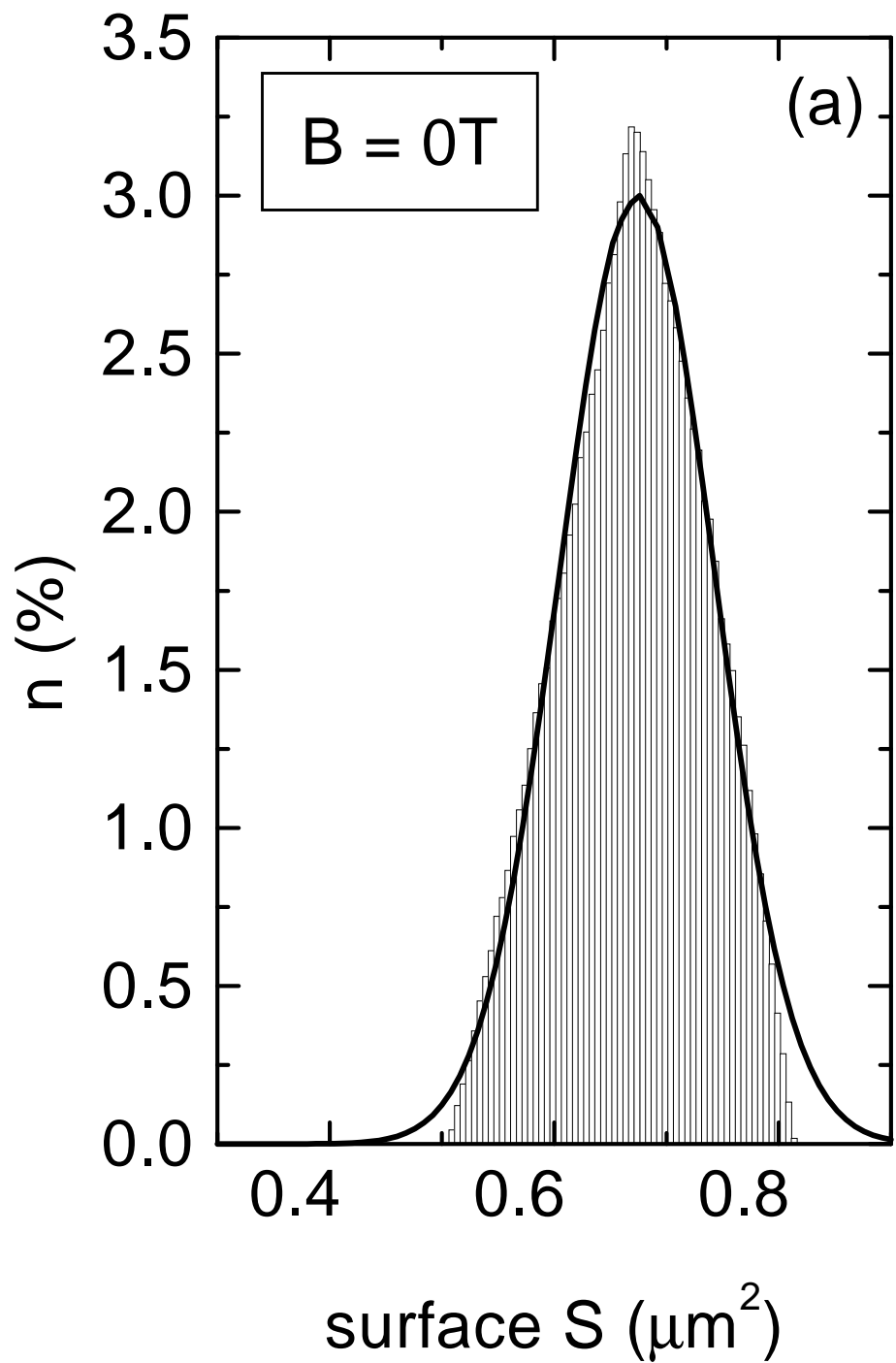
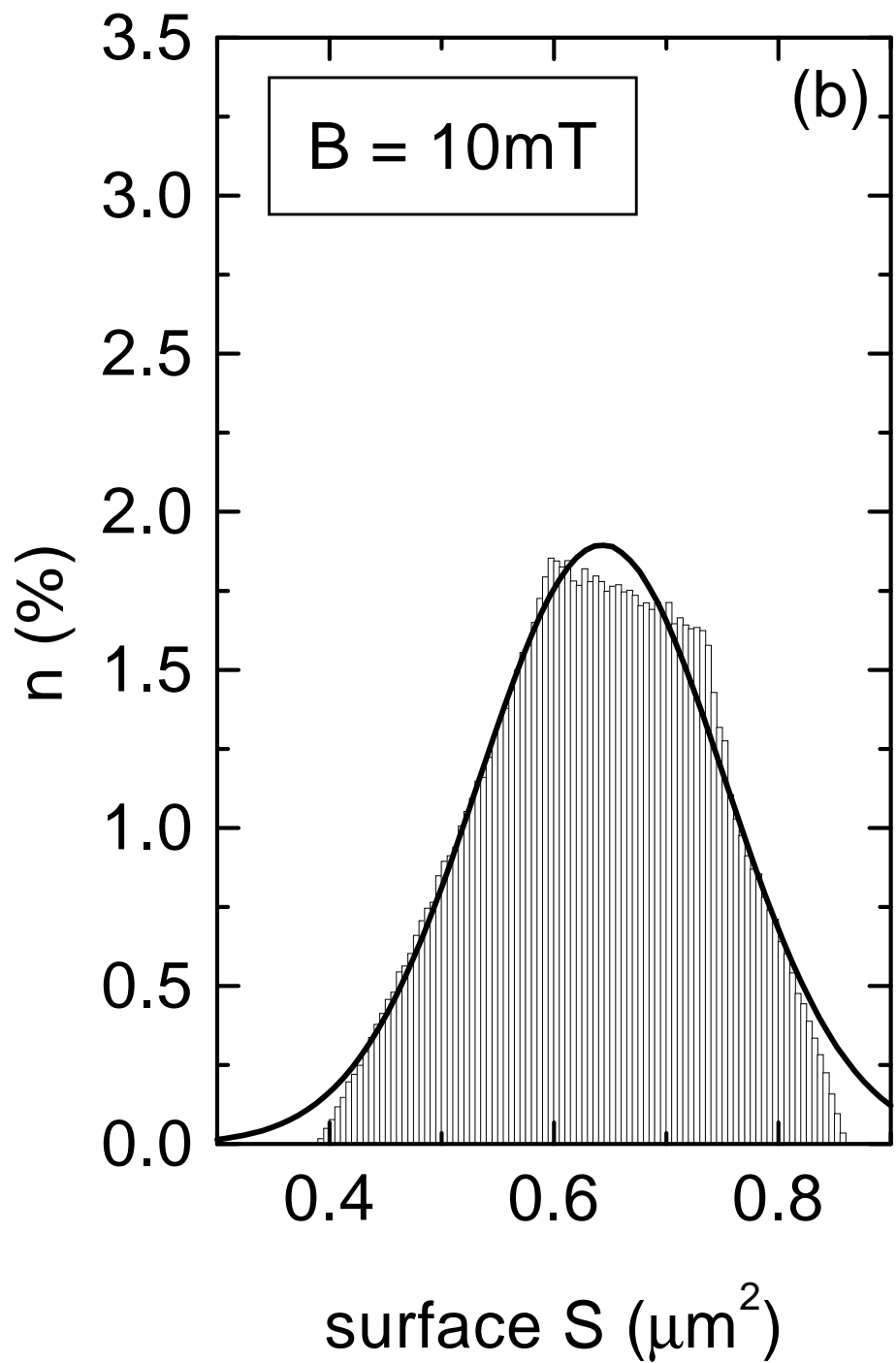
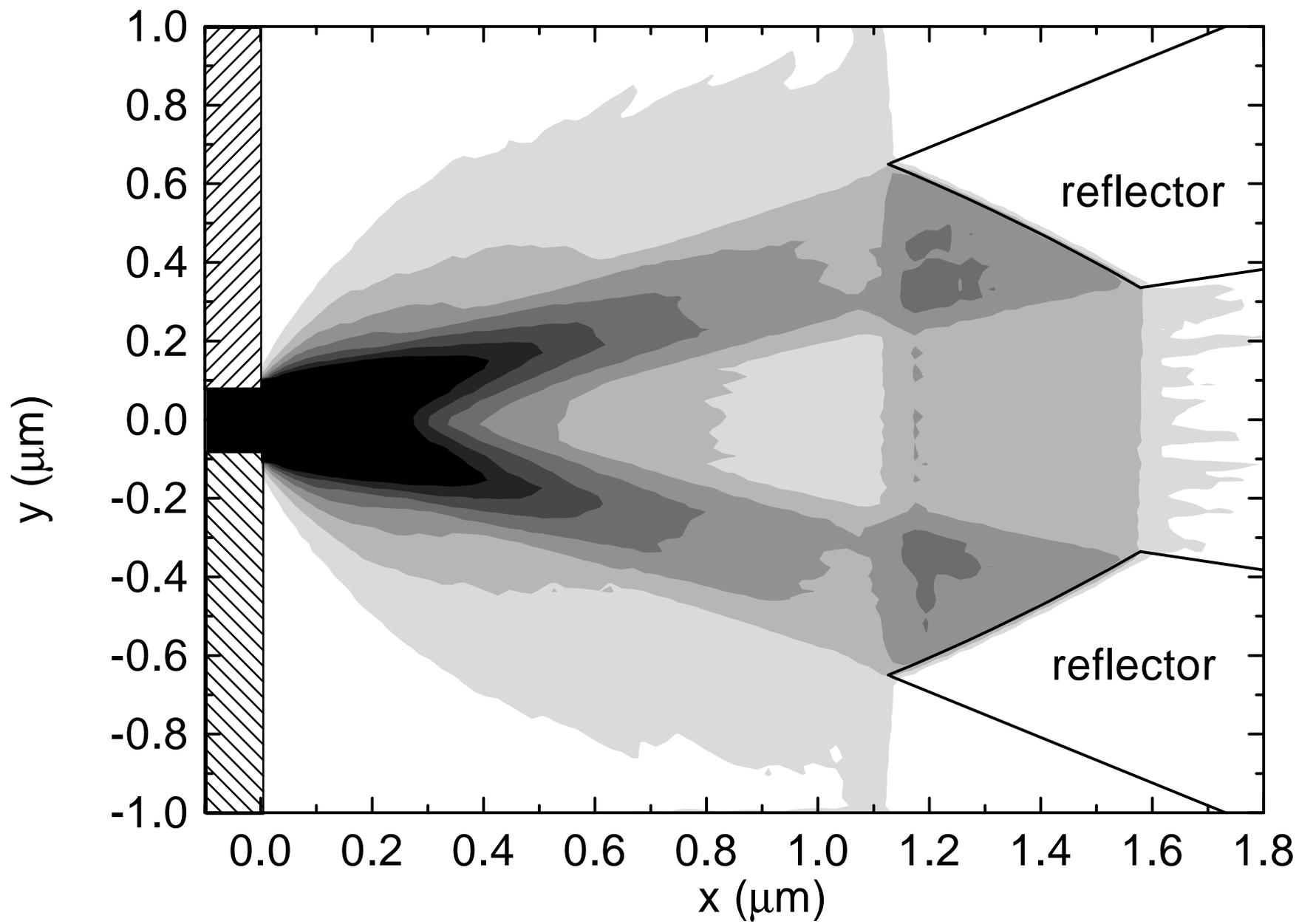
\end{document}